# ELECTRODYNAMICAL FORBIDDANCE OF STRONG QUADRUPOLE LIGHT-MOLECULE INTERACTION IN THE METHANE MOLECULE AND ITS MANIFESTATION IN SURFACE ENHANCED HYPER RAMAN SCATTERING SPECTRA

**[1]V.P. Chelibanov, [2]A.M. Polubotko**

[1]State University of Information Technologies, Mechanics and Optics, Kronverkskii 49, 197101 Saint Petersburg, RUSSIA  E-mail: Chelibanov@gmail.com

[2] A.F. Ioffe Physico-Technical Institute, Politechnicheskaya 26, 194021 Saint Petersburg, RUSSIA E-mail: alex.marina@mail.ioffe.ru

## ABSTRACT

It is demonstrated in the framework of the Dipole-Quadrupole theory, that strong quadrupole light-molecule interaction, which is responsible for the most enhancement of SEHRS in the methane molecule, which belongs to the $T_d$ symmetry group, experiences so-called electrodynamical forbiddance due to electrodynamical law $divE = 0$, and does not influence on the formation of the SEHRS spectra. This forbiddance results in the fact that the lines, caused by the totally symmetric vibrations, transforming after the unit irreducible representation, which are observed in symmetrical molecules with another sufficiently high groups of the point symmetry, such as pyrazine and phenazine, with $D_{2h}$ symmetry group, must be slight, or be absent at all. In this case in methane the most enhanced lines are those, caused by vibrations, transforming by the irreducible representation $T_1$, or $T_2$.

## INTRODUCTION

The theory of surface enhance optical processes is a strongly developing area of science and is of great interest. A large progress in this area was achieved after developing of the Dipole-Quadrupole theory, which allows to explain a lot of experimental facts in such phenomena as SERS [1], surface enhanced infrared absorption [2-4] and surface enhanced Hyper Raman scattering [5,6]. The most important feature of this theory is possibility of explanation of forbidden lines in molecules with sufficiently high symmetry. Such line appear in experimental spectra in all pointed processes. In particular in SEHRS they appear in such molecules as pyrazine and phenazine [7-9] and their appearance was successfully explained in [5,6,10]. In this work we shall demonstrate that in the methane molecule, which refer to the $T_d$ symmetry group, the lines, caused by totally symmetric vibrations, transforming after the unit irreducible representation, which manifest strongly in the pyrazine and phenazine molecules either very weak, either absent at all due to so-called electrodynamical forbiddance, which is sequence of the electrodynamical law

$div\mathbf{E} = 0$

## ELECTRODYNAMICAL FORBIDDANCE AND PECULIARITIES OF THE SEHRS SPECTRA OF THE METHANE MOLECULE

The SEHRS cross-section of symmetrical molecule in the framework of the Dipole-Quadrupole theory can be obtain with the quantum mechanical perturbation theory [11] and is expressed via

$$\frac{d}{dt}\left|w^{(3)}_{(n,\overline{V}\pm 1),(n,\overline{V})}(t)\right|^2$$

where $W^{(3)}$ is the third member of the expansion of the coefficient of the perturbation wavefunction with quantum numbers $(n,\overline{V}\pm 1)$. Here $n$ is the quantum number of the ground state of the molecule, $\overline{V}$ designates the totality of vibrational quantum numbers $(V_1, V_2 ... V_s ...)$. The expression $(\overline{V}\pm 1)$ designates a vibrational quantum state, when one of the quantum numbers changes on one unit. The general expression for $W^{(3)}$ has the form.

$$w^{(3)}_{(n,\overline{V}\pm 1),(n,\overline{V})}(t) = \left(-\frac{i}{\hbar}\right)^3 \sum_{\substack{m \\ m\neq n}} \Bigg[ \int_0^t \langle n,\overline{V}\pm 1|\hat{\mathbf{H}}^{inc}_{e-r}+\hat{\mathbf{H}}^{scat}_{e-r}|k,\overline{V}\pm 1\rangle dt_1 \times$$

$$\times \int_0^{t_1}\langle k,\overline{V}\pm 1|\hat{\mathbf{H}}^{inc}_{e-r}+\hat{\mathbf{H}}^{scat}_{e-r}|m,\overline{V}\pm 1\rangle dt_2 \times \int_0^{t_2}\langle m,\overline{V}\pm 1|\hat{\mathbf{H}}^{inc}_{e-r}+\hat{\mathbf{H}}^{scat}_{e-r}|n,\overline{V}\rangle dt_3 +$$

$$+ \int_0^t \langle n,\overline{V}\pm 1|\hat{\mathbf{H}}^{inc}_{e-r}+\hat{\mathbf{H}}^{scat}_{e-r}|k,\overline{V}\pm 1\rangle dt_1 \times \times \int_0^{t_1}\langle k,\overline{V}\pm 1|\hat{\mathbf{H}}^{inc}_{e-r}+\hat{\mathbf{H}}^{scat}_{e-r}|m,\overline{V}\rangle dt_2$$

$$\times \int_0^{t_2}\langle m,\overline{V}|\hat{\mathbf{H}}^{inc}_{e-r}+\hat{\mathbf{H}}^{scat}_{e-r}|n,\overline{V}\rangle dt_3 + + \int_0^t \langle n,\overline{V}\pm 1|\hat{\mathbf{H}}^{inc}_{e-r}+\hat{\mathbf{H}}^{scat}_{e-r}|k,\overline{V}\rangle dt_1 \times$$

$$\times \int_0^{t_1}\langle k,\overline{V}|\hat{\mathbf{H}}^{inc}_{e-r}+\hat{\mathbf{H}}^{scat}_{e-r}|m,\overline{V}\rangle dt_2 \times \int_0^{t_2}\langle m,\overline{V}|\hat{\mathbf{H}}^{inc}_{e-r}+\hat{\mathbf{H}}^{scat}_{e-r}|n,\overline{V}\rangle dt_3 \Bigg].$$

Here $\hat{\mathbf{H}}^{inc}_{e-r}$ и $\hat{\mathbf{H}}^{scat}_{e-r}$ are Hamiltonians of light-molecule interaction for the incident and scattered fields, which have a form

$$\hat{\mathbf{H}}^{inc}_{e-r} = |\mathbf{E}_{inc}|\frac{(\mathbf{e}^*\mathbf{f}_e^*)_{inc}e^{i\omega_{inc}t}+(\mathbf{ef}_e)_{inc}e^{-i\omega_{inc}t}}{2}$$

$$\hat{\mathbf{H}}^{scat}_{e-r} = |\mathbf{E}_{scat}|\frac{(\mathbf{e}^*\mathbf{f}_e^*)_{scat}e^{i\omega_{scat}t}+(\mathbf{ef}_e)_{scat}e^{-i\omega_{scat}t}}{2}$$

$\mathbf{E}_{inc}$ and $\mathbf{E}_{scat}$ are the vectors of the incident and scattered fields? $\omega_{inc}$ and $\omega_{scat}$ are corresponding frequencies, $\mathbf{e}$ designates the polarization vector of the corresponding fields.

$$f_{e\alpha} = d_{e\alpha} + \frac{1}{2E_\alpha}\sum_\beta \frac{\partial E_\alpha}{\partial x_\beta} Q_{e\alpha\beta}$$

is an $\alpha$ component of the generalized vector of interaction of light with molecules.

$$d_{e\alpha} = \sum_i e x_{i\alpha}$$

$$Q_{e\alpha\beta} = \sum_i e x_{i\alpha} x_{i\beta}$$

is an $\alpha$ component of the dipole moment vector and $\alpha\beta$ component of the quadrupole moment tensor of electrons. Here under $x_{i\alpha}$ and $x_{i\beta}$ we mean the coordinates $x, y, z$ of the $i$ electron.

In the methane molecule the dipole and quadrupole moments with various indices transform after irreducible representation of the symmetry group. However the quadrupole moments with various indices transform after reducible representations. Therefore it is convenient to express the later values via linear combinations of these moments $Q_1, Q_2$ and $Q_3$, which transform after irreducible representations.. In the $T_d$ group these combinations have the form

$$Q_1 = \frac{1}{3}(Q_{xx} + Q_{yy} + Q_{zz})$$

$$Q_2 = \frac{1}{2}(Q_{xx} - Q_{yy})$$

$$Q_3 = \frac{1}{4}(Q_{xx} + Q_{yy} - 2Q_{zz}) /$$

and

$$Q_{xx} = Q_1 + \frac{2}{3}Q_3 + Q_2$$

$$Q_{yy} = Q_1 + \frac{2}{3}Q_3 - Q_2$$

$$Q_{zz} = Q_1 + \frac{2}{3}Q_3 - 2Q_2$$

Then one can obtain the following expression for $|\mathbf{E}|(\mathbf{ef}_e)$, which is contained in the light-molecule interaction Hamiltonian

$$|\mathbf{E}|(\mathbf{ef}_e) = (\mathbf{Ed}) + \frac{1}{2} div\mathbf{E} \times \left(Q_1 + \frac{2}{3}Q_3\right) + \frac{1}{2}\left(\frac{\partial E_x}{\partial x} - \frac{\partial E_y}{\partial y} - 2\frac{\partial E_z}{\partial z}\right)Q_2$$

$$+ \frac{1}{2} \sum_{\substack{\alpha\beta \\ \alpha \neq \beta}} \frac{\partial E_\alpha}{\partial x_\beta} Q_{e\alpha\beta} \tag{1}$$

In accordance with our previous ideas [1,5,6,12] the value $Q_1$ is the main moment with a constant sign, which transform after the unit irreducible representation and is responsible for the most enhancement. In particular in the pyrazine and phenazine molecules the main quadrupole moments, which has another form $Q_{main} = (Q_{xx}, Q_{yy}, Q_{zz})$ are responsible for appearance of forbidden lines in the SEHRS spectra [5,6,10], caused by vibrations, transforming after the unit irreducible representation. In our case one can see from (1) that there is a factor $div\mathbf{E} = \mathbf{0}$, which make equal to zero the contributions from this moment. This result is an Electrodynamical forbiddance of strong quadrupole light-molecule interaction, which realizes in the methane molecule. Due to this forbiddance, associated with the cubic symmetry and with electrodynamical law, all terms, associated with $Q_1$ and $Q_3$ are equal to zero and do not influence on formation of the enhanced SEHRS spectra. The expression for the surface enhanced Hyper Raman spectra of symmetrical molecule, obtained in the framework of adiabatic perturbation theory [12,13] has the form

$$d\sigma_{SEHRS_s}\begin{pmatrix} St \\ AnSt \end{pmatrix} = \frac{\omega_{inc}\omega_{scat}^3}{64\pi^2\hbar^4\varepsilon_0^2 c^4}|\mathbf{E}_{inc}|_{vol}^2 \frac{|\mathbf{E}_{inc}|_{surf}^2 |\mathbf{E}_{inc}|_{surf}^2 |\mathbf{E}_{scat}|_{surf}^2}{|\mathbf{E}_{inc}|_{vol}^2 |\mathbf{E}_{inc}|_{vol}^2 |\mathbf{E}_{scat}|_{vol}^2} \times$$

$$\times \sum_p \begin{pmatrix} \frac{V_{(s,p)}+1}{2} \\ \frac{V_{(s,p)}}{2} \end{pmatrix} \left| \sum_{f_1, f_2, f_3} T_{(s,p),(f_1-f_2-f_3)} \right|^2 dO$$

where the signs *surf* and *vol* near the fields designate that the incident and the scattered fields correspond to the surface field and to the field in a free space respectively. $V_{(s,p)}$ is a vibrational quantum number of the $(s,p)$ vibrational mode. $s$ numerates the groups of degenerate states, while $p$ the states inside the group. Under $f_1, f_2$ and $f_3$ we mean the dipole and quadrupole moments, or their linear combinations, transforming after irreducible representations of the symmetry group of the molecule, $T_{(s,p),(f_1-f_2-f_3)}$ are scattering contributions, caused by the scattering on the moments $f_1, f_2$ and $f_3$, which further we shall designate as $(f_1 - f_2 - f_3)$. Another designations are conventional. Detailed expressions for the contributions are listed in Appendix. As it is demonstrated in [5,6] the contributions in the cross-section $T_{(s,p),(f_1-f_2-f_3)}$ obey selection rules,

$$\Gamma_{(s,p)} \in \Gamma_{f_1}\Gamma_{f_2}\Gamma_{f_3},$$

where $\Gamma$ designates irreducible representation of the $(s,p)$ vibrational mode and of the $f_1, f_2$ and $f_3$ moments respectively. As it was pointed out above all the moments can be classified as the main ones, which are responsible for the strong enhancement and the minor ones with sufficiently slight scattering. In molecules, adsorbed on strongly rough surface of silver, gold and copper, the main moments are $Q_{main} = (Q_{xx}, Q_{yy}, Q_{zz})$ and their linear combinations with constant sign. Due to the enhancement of the component $E_z$, which is perpendicular to the surface, the moment $d_z$ refer to the main moments too. Another moments $Q_{xy}, Q_{xz}, Q_{yz}$ and also linear combinations of the moments $Q_{xx}, Q_{yy}$ and $Q_{zz}$ with a changeable sign, transforming after irreducible representations of the symmetry group are not responsible for the strong enhancement and therefore one refer them to the minor moments. For monolayer coverage, because the components $E_x$ and $E_y$ which are parallel to the surface tend to zero, the moments $d_x$ and $d_y$ refer to the minor moments too. For multilayer coverage, because of orientation of molecules with respect to the surface can be arbitrary in upper layers, the moments $d_x$ and $d_y$ also can take part in the enhancement and be responsible for appearance of some lines. Therefore, in this case these moments can be refer to the main moments. However it is necessary to take into account that the enhancement in the second and upper layers strongly decrease due to so-called the "first layer effect". Therefore in spite these moments take part in the enhancement, however their influence is not strong.

Let us consider the case of monolayer coverage in detail. Then all the contributions can be classified in accordance with the enhancement degree, in accordance with the moments classification.
1. The most enhanced contributions are the ones $(Q_{main} - Q_{main} - Q_{main})$.
2. The contributions $(Q_{main} - Q_{main} - d_z)$ are enhanced also, but with a lesser degree than the previous ones.
3. The contributions of $(Q_{main} - d_z - d_z)$ type also must be enhanced, but with a lesser degree, than the previous two.
4. The contributions $(d_z - d_z - d_z)$ also will be enhanced, but with the lesser degree than the previous three.

Another contributions, which contain minor dipole and quadrupole moments and therefore they must be enhanced significantly smaller, than the previous four. However in the methane molecule all the contributions, which contain the main quadrupole moment $Q_{main} = Q_1$ are forbidden due to the Electrodynamical forbiddance. Therefore the first three contributions in this molecule are equal to zero and the most enhanced contribution is $(d_z - d_z - d_z)$. The theoretical group analysis demonstrates, that this contribution transforms after reducible representation, which contains two irreducible representations $F_1$ and $F_2$. Thus the most enhanced lines do not belong to the vibrations with the unit irreducible representation, as it was in molecules with lower symmetry, $D_{2h}$ in particular. Designating minor dipole and quadrupole moments as $Q_{minor}$ and $d_{minor}$, the remaining contributions, which may be enhanced due to the presence of the $d_z$ moment, $(d_z - d_z - Q_{minor})$, $(d_z - d_z - d_{minor})$, $(d_z - d_{minor} - d_{minor})$ can contribute in principle in the lines, caused by the totally symmetric vibrations. However these contributions are significantly weaker than the contribution $(d_z - d_z - d_z)$. Therefore these lines must be very weak, or must be absent at all in the SEHRS spectra of methane.

One should note, that for the multilayer coverage, the moments $d_x$ and $d_y$ also can refer to the main moments. However as it has been pointed above, because the enhancement in the first and upper layers is significantly weaker, the contributions $(d_z - d_x - d_x)$, $(d_z - d_y - d_y)$ and $(d_z - d_x - d_y)$ will be enhanced significantly lower, than the contribution $(d_z - d_z - d_z)$. Therefore even the multilayer coverage must not result in appearance of strong lines, caused by totally symmetric vibrations. Therefore the lines caused by vibrations with $F_1$ and $F_2$ irreducible representations must dominate in the spectrum.

The Electrodynamical forbiddance of the strong quadrupole light-molecule interaction must be observed in molecules, belonging to another cubic symmetry groups $O, O_h, T$, and $T_h$, where linear combinations of the quadrupole moments $Q_1, Q_2$ and $Q_3$ have the same form. This must result in the fact, that in another molecules, such as in methane, the lines, caused by the totally symmetric vibrations must not experience the most enhancement. The most enhanced lines will be caused by another irreducible representations. The matter of presence of the lines with totally symmetric vibrations in the spectra of these molecules is very difficult and is associated with specific symmetry group of these molecules. From our point of view, this matter must be solved together with experimental investigations of the SEHRS spectra of specific molecules.

# APPENDIX

## GENERAL FORM FOR THE SCATTERING CONTRIBUTIONS IN MOLECULES WITH SYMMETRY GROUPS $O, O_h, T, T_d$ AND $T_h$

Below we shall write out the expressions for the sum of the contributions? Caused by combinations of moments of the same type. For example by the combinations of the $(d-d-d)$ type. Here we mean all possible combinations of the moments $d_\alpha$, $\alpha = (x, y, z)$.

The expressions for the contributions are very cumbersome, which can be strongly simplified in case one introduce the scattering tensor.

$$C_{V_{(s,p)}}[f_1, f_2, f_3]\begin{pmatrix} St \\ AnSt \end{pmatrix} =$$

$$\sum_{m,r,l \neq n} \frac{\langle n|f_3|m\rangle\langle m|f_2|r\rangle\langle r|f_1|l\rangle R_{n,l,(s,p)}}{(E_n^{(0)} - E_l^{(0)})(\omega_{m,n} - 2\omega_{inc})(\omega_{r,n} \pm \omega_{(s,p)} - \omega_{inc})} +$$

$$+ \sum_{m,r,l \neq n} \frac{\langle n|f_2|m\rangle\langle m|f_3|r\rangle\langle r|f_1|l\rangle R_{n,l,(s,p)}}{(E_n^{(0)} - E_l^{(0)})(\omega_{m,n} + \omega_{scat} - \omega_{inc})(\omega_{r,n} \pm \omega_{(s,p)} - \omega_{inc})} +$$

$$+ \sum_{m,r,l \neq n} \frac{\langle n|f_2|m\rangle\langle m|f_1|r\rangle\langle r|f_3|l\rangle R_{n,l,(s,p)}}{(E_n^{(0)} - E_l^{(0)})(\omega_{m,n} + \omega_{scat} - \omega_{inc})(\omega_{r,n} \pm \omega_{(s,p)} + \omega_{scat})} +$$

$$+ \sum_{m,r,l \neq n} \frac{\langle n|f_3|m\rangle\langle m|f_2|r\rangle R^*_{r,l,(s,p)}\langle l|f_1|n\rangle}{(E_r^{(0)} - E_l^{(0)})(\omega_{m,n} - 2\omega_{inc})(\omega_{r,n} \mp \omega_{(s,p)} - \omega_{inc})} +$$

$$+ \sum_{m,r,l \neq n} \frac{\langle n|f_2|m\rangle\langle m|f_3|r\rangle R^*_{r,l,(s,p)}\langle l|f_1|n\rangle}{(E_r^{(0)} - E_l^{(0)})(\omega_{m,n} + \omega_{scat} - \omega_{inc})(\omega_{r,n} \mp \omega_{(s,p)} - \omega_{inc})} +$$

$$+ \sum_{m,r,l \neq n} \frac{\langle n|f_2|m\rangle\langle m|f_1|r\rangle R^*_{r,l,(s,p)}\langle l|f_3|n\rangle}{(E_r^{(0)} - E_l^{(0)})(\omega_{m,n} + \omega_{scat} - \omega_{inc})(\omega_{r,n} \mp \omega_{(s,p)} + \omega_{scat})} +$$

$$+ \sum_{m,r,l \neq n} \frac{\langle n|f_3|m\rangle\langle m|f_2|l\rangle R_{r,l,(s,p)}\langle r|f_1|n\rangle}{(E_r^{(0)} - E_l^{(0)})(\omega_{m,n} \pm \omega_{(s,p)} - 2\omega_{inc})(\omega_{r,n} - \omega_{inc})} +$$

$$+ \sum_{m,r,l \neq n} \frac{\langle n|f_2|m\rangle\langle m|f_3|l\rangle R_{r,l,(s,p)}\langle r|f_1|n\rangle}{(E_r^{(0)} - E_l^{(0)})(\omega_{m,n} + \omega_{scat} \pm \omega_{(s,p)} - \omega_{inc})(\omega_{r,n} - \omega_{inc})} +$$

$$+ \sum_{m,r,l \neq n} \frac{\langle n|f_2|m\rangle\langle m|f_1|l\rangle R_{r,l,(s,p)}\langle r|f_3|n\rangle}{(E_r^{(0)} - E_l^{(0)})(\omega_{m,n} + \omega_{scat} \pm \omega_{(s,p)} - \omega_{inc})(\omega_{r,n} + \omega_{scat})} +$$

$$+ \sum_{m,r,l \neq n} \frac{\langle n|f_3|m\rangle R^*_{m,l,(s,p)}\langle l|f_2|r\rangle\langle r|f_1|n\rangle}{(E_m^{(0)} - E_l^{(0)})(\omega_{m,n} \mp \omega_{(s,p)} - 2\omega_{inc})(\omega_{r,n} - \omega_{inc})} +$$

$$+ \sum_{m,r,l \neq n} \frac{\langle n|f_2|m\rangle R^*_{m,l,(s,p)}\langle l|f_3|r\rangle\langle r|f_1|n\rangle}{(E_m^0 - E_l^0)(\omega_{m,n} + \omega_{scat} \mp \omega_{(s,p)} - \omega_{inc})(\omega_{r,n} - \omega_{inc})} +$$

$$+ \sum_{m,r,l \neq n} \frac{\langle n|f_2|m\rangle R^*_{m,l,(s,p)}\langle l|f_1|r\rangle\langle r|f_3|n\rangle}{(E_m^{(0)} - E_l^{(0)})(\omega_{m,n} + \omega_{scat} \mp \omega_{(s,p)} - \omega_{inc})(\omega_{r,n} + \omega_{scat})} +$$

$$+ \sum_{m,r,l \neq n} \frac{\langle n|f_3|l\rangle R_{m,l,(s,p)}\langle m|f_2|r\rangle\langle r|f_1|n\rangle}{(E_m^{(0)} - E_l^{(0)})(\omega_{m,n} - 2\omega_{inc})(\omega_{r,n} - \omega_{inc})} +$$

$$+ \sum_{m,r,l \neq n} \frac{\langle n|f_2|l\rangle R_{m,l,(s,p)}\langle m|f_3|r\rangle\langle r|f_1|n\rangle}{(E_m^{(0)} - E_l^{(0)})(\omega_{m,n} + \omega_{scat} - \omega_{inc})(\omega_{r,n} - \omega_{inc})} +$$

$$+ \sum_{m,r,l \neq n} \frac{\langle n|f_2|l\rangle R_{m,l,(s,p)}\langle m|f_1|r\rangle\langle r|f_1|n\rangle}{(E_m^{(0)} - E_l^{(0)})(\omega_{m,n} + \omega_{scat} - \omega_{inc})(\omega_{r,n} + \omega_{scat})} +$$

$$+ \sum_{m,r,l \neq n} \frac{R^*_{n,l(s,p)}\langle l|f_3|m\rangle\langle m|f_2|r\rangle\langle r|f_1|n\rangle}{(E_n^{(0)} - E_l^{(0)})(\omega_{m,n} - 2\omega_{inc})(\omega_{r,n} - \omega_{inc})} +$$

$$+ \sum_{m,r,l \neq n} \frac{R^*_{n,l,(s,p)}\langle l|f_2|m\rangle\langle m|f_3|r\rangle\langle r|f_1|n\rangle}{(E_n^{(0)} - E_l^{(0)})(\omega_{m,n} + \omega_{scat} - \omega_{inc})(\omega_{r,n} - \omega_{inc})} +$$

$$+ \sum_{m,r,l \neq n} \frac{R^*_{n,l,(s,p)}\langle l|f_2|m\rangle\langle m|f_1|r\rangle\langle r|f_3|n\rangle}{(E_n^{(0)} - E_l^{(0)})(\omega_{m,n} + \omega_{scat} - \omega_{inc})(\omega_{r,n} + \omega_{scat})}$$

In accordance with [1,6], $E_n^{(0)}$ designates the energy of the ground electronic state of the molecule, $E_l^{(0)}, E_r^{(0)}, E_m^{(0)}$ the energies of the excited state of the molecule with motionless nuclei. $R_{n,l,(s,p)}$ are the coefficients of excitation of the state $l$ from the ground state $n$, by the $(s,p)$ vibrational mode. More detailed expression for these coefficients one can find in [1]. The coefficients $R_{r,l,(s,p)}$ and $R_{m,l,(s,p)}$ are the same, but only for the excited states $r$ and m. $\omega_{(s,p)}$ is a vibrational frequency of the $(s,p)$ vibrational mode. The expressions $\omega_{m,n}$ are

$$\omega_{m,n} = \frac{E_m^{(0)} - E_n^{(0)}}{\hbar}$$

Let us point out the following property of the scattering tensor

$$C_{V_{(s,p)}}[f_i, f_j, (a_1 f_k + a_2 f_m)] = a_1 C_{V_{(s,p)}}[f_i, f_j, f_k] + a_2 C_{V_{(s,p)}}[f_i, f_j, f_m]$$
$$C_{V_{(s,p)}}[f_i, (a_1 f_j + a_2 f_k), f_m] = a_1 C_{V_{(s,p)}}[f_i, f_j, f_m] + a_2 C_{V_{(s,p)}}[f_i, f_k, f_m]$$
$$C_{V_{(s,p)}}[(a_1 f_i + a_2 f_j), f_k, f_m] = a_1 C_{V_{(s,p)}}[f_i, f_k, f_m] + a_2 C_{V_{(s,p)}}[f_j, f_k, f_m]$$

Using this property, taking into account (1), the expressions for the sums of all contributions $T_{f_1-f_2-f_3}$ of the same type, $T_{d-d-d}$ for example will have the form

$$T_{d-d-d} = \sum_{i,j,k} C_{V_{(s,p)}}[d_i, d_j, d_k] (e^*_{scat,i} e_{inc,j} e_{inc,k})_{surf} \tag{2}$$

Further

$$T_{d-d-Q} = \sum_{\substack{i,j,\chi,\eta \\ \chi \neq \eta}} C_{V_{(s,p)}}[d_i, d_j, Q_{\chi\eta}] \left( e^*_{scat,i} e_{inc,j} \frac{1}{2|\mathbf{E}_{inc}|} \frac{\partial E^{inc}_\chi}{\partial x_\eta} \right)_{surf}$$

$$+ \sum_{i,j} C_{V_{(s,p)}}[d_i, d_j, Q_2] \left( e^*_{scat,i} e_{inc,j} \frac{1}{2|\mathbf{E}_{inc}|} \left( \frac{\partial E^{inc}_x}{\partial x} - \frac{\partial E^{inc}_y}{\partial y} - 2\frac{\partial E^{inc}_z}{\partial z} \right) \right)_{surf} \tag{3}$$

$$T_{d-Q-d} = \sum_{\substack{i,\gamma,\delta,k \\ \gamma \neq \delta}} C_{V_{(s,p)}}[d_i, Q_{\gamma\delta}, d_k] \left( e^*_{scat,i} \frac{1}{2|\mathbf{E}_{inc}|} \frac{\partial E^{inc}_\gamma}{\partial x_\delta} e_{inc,k} \right)_{surf} +$$

$$T_{d-Q-Q} = \sum_{\substack{i,\gamma,\delta,\chi,\eta \\ \gamma \neq \delta \\ \chi \neq \eta}} C_{V_{(s,p)}}[d_i, Q_{\gamma\delta}, Q_{\chi\eta}] \left( e^*_{scat,i} \frac{1}{2|\mathbf{E}_{inc}|} \frac{\partial E^{inc}_\gamma}{\partial x_\delta} \frac{1}{2|\overline{E}_{inc}|} \frac{\partial E^{inc}_\chi}{\partial x_\eta} \right)_{surf} +$$

$$+ \sum_{\substack{i,\gamma,\delta \\ \gamma \neq \delta}} C_{V_{(s,p)}} \left[ d_i, Q_{\gamma\delta}, Q_2 \right] \left( e_{scat,i}^* \frac{1}{2|\mathbf{E}_{inc}|} \frac{\partial E_\gamma^{inc}}{\partial x_\delta} \frac{1}{2|\mathbf{E}_{inc}|} \left( \frac{\partial E_x^{inc}}{\partial x} - \frac{\partial E_y^{inc}}{\partial y} - 2\frac{\partial E_z^{inc}}{\partial z} \right) \right)_{surf} +$$

$$+ \sum_{\substack{i,\chi,\eta \\ \chi \neq \eta}} C_{V_{(s,p)}} \left[ d_i, Q_2, Q_{\chi\eta} \right] \left( e_{scat,i}^* \frac{1}{2|\mathbf{E}_{inc}|} \left( \frac{\partial E_x^{inc}}{\partial x} - \frac{\partial E_y^{inc}}{\partial y} - 2\frac{\partial E_z^{inc}}{\partial z} \right) \frac{1}{2|\overline{E}_{inc}|} \frac{\partial E_\chi^{inc}}{\partial x_\eta} \right)_{surf} +$$

(5)

$$T_{Q-d-Q} = \sum_{\substack{\alpha,\beta,j,\gamma,\delta \\ \alpha \neq \beta \\ \gamma \neq \delta}} C_{V_{(s,p)}} \left[ Q_{\alpha\beta}, d_j, Q_{\gamma\delta} \right] \left( \frac{1}{2|\mathbf{E}_{inc}|} \frac{\partial E_\alpha^{scat^*}}{\partial x_\beta} e_{inc,j} \frac{1}{2|\mathbf{E}_{inc}|} \frac{\partial E_\gamma^{inc}}{\partial x_\delta} \right)_{surf} +$$

$$+ \sum_{\substack{\alpha,\beta,j \\ \alpha \neq \beta}} C_{V_{(s,p)}} \left[ Q_{\alpha\beta}, d_j, Q_2 \right] \left( \frac{1}{2|\mathbf{E}_{inc}|} \frac{\partial E_\alpha^{scat^*}}{\partial x_\beta} e_{inc,j} \frac{1}{2|\mathbf{E}_{inc}|} \left( \frac{\partial E_x^{inc}}{\partial x} - \frac{\partial E_y^{inc}}{\partial y} - 2\frac{\partial E_z^{inc}}{\partial z} \right) \right)_{surf} +$$

$$\sum_{\substack{j,\chi,\eta \\ \chi \neq \eta}} C_{V_{(s,p)}} \left[ Q_2, d_j, Q_{\chi\eta} \right] \left( \frac{1}{2|\mathbf{E}_{inc}|} \left( \frac{\partial E_x^{scat^*}}{\partial x} - \frac{\partial E_y^{scat^*}}{\partial y} - 2\frac{\partial E_z^{scat^*}}{\partial z} \right) e_{inc,j} \frac{1}{2|\mathbf{E}_{inc}|} \frac{\partial E_\chi^{inc}}{\partial x_\eta} \right)_{surf} +$$

$$+ \sum_j C_{V_{(s,p)}} \left[ Q_2, d_j, Q_2 \right] \left( \begin{array}{l} \frac{1}{2|\mathbf{E}_{inc}|} \left( \frac{\partial E_x^{scat^*}}{\partial x} - \frac{\partial E_y^{scat^*}}{\partial y} - 2\frac{\partial E_z^{scat^*}}{\partial z} \right) e_{inc,j} \times \\ \times \frac{1}{2|\mathbf{E}_{inc}|} \left( \frac{\partial E_x^{inc}}{\partial x} - \frac{\partial E_y^{inc}}{\partial y} - 2\frac{\partial E_z^{inc}}{\partial z} \right) \end{array} \right)_{surf}$$

(6)

$$T_{Q-Q-d} = \sum_{\substack{\alpha,\beta,\gamma,\delta,k \\ \alpha \neq \beta \\ \gamma \neq \delta}} C_{V_{(s,p)}} \left[ Q_{\alpha\beta}, Q_{\gamma\delta}, d_k \right] \left( \frac{1}{2|\mathbf{E}_{inc}|} \frac{\partial E_\alpha^{scat^*}}{\partial x_\beta} \frac{1}{2|\mathbf{E}_{inc}|} \frac{\partial E_\gamma^{inc}}{\partial x_\delta} e_{inc,k} \right)_{surf} +$$

$$+ \sum_{\substack{\alpha,\beta,k \\ \alpha \neq \beta}} C_{V_{(s,p)}} \left[ Q_{\alpha\beta}, Q_2, d_k \right] \left( \frac{1}{2|\mathbf{E}_{inc}|} \frac{\partial E_\alpha^{scat^*}}{\partial x_\beta} \frac{1}{2|\mathbf{E}_{inc}|} \left( \frac{\partial E_x^{inc}}{\partial x} - \frac{\partial E_y^{inc}}{\partial y} - 2\frac{\partial E_z^{inc}}{\partial z} \right) e_{inc,k} \right)_{surf} +$$

$$+ \sum_{\substack{\gamma,\delta,k \\ \gamma\neq\delta}} C_{V_{(s,p)}}\left[Q_2, Q_{\gamma\delta}, d_k\right]\left(\frac{1}{2|\mathbf{E}_{inc}|}\left(\frac{\partial E_x^{scat^*}}{\partial x} - \frac{\partial E_y^{scat^*}}{\partial y} - 2\frac{\partial E_z^{scat^*}}{\partial z}\right)\frac{1}{2|\mathbf{E}_{inc}|}\frac{\partial E_\gamma^{inc}}{\partial x_\delta}e_{inc,k}\right)_{surf} +$$

(7)

$$T_{Q-Q-Q} = \sum_{\substack{\alpha,\beta,\gamma,\delta,\chi,\eta \\ \alpha\neq\beta \\ \gamma\neq\delta \\ \chi\neq\eta}} C_{V_{(s,p)}}\left[Q_{\alpha\beta}, Q_{\gamma\delta}, Q_{\chi\eta}\right]\left(\frac{1}{2|\mathbf{E}_{scat}|}\frac{\partial E_\alpha^{scat^*}}{\partial x_\beta}\frac{1}{2|\mathbf{E}_{inc}|}\frac{\partial E_\gamma^{inc}}{\partial x_\delta}\frac{1}{2|\mathbf{E}_{inc}|}\frac{\partial E_\chi^{inc}}{\partial x_\eta}\right)_{surf} +$$

$$+ \sum_{\substack{\alpha,\beta,\gamma,\delta \\ \alpha\neq\beta \\ \gamma\neq\delta}} C_{V_{(s,p)}}\left[Q_{\alpha\beta}, Q_{\gamma\delta}, Q_2\right]\left(\begin{array}{c}\frac{1}{2|\mathbf{E}_{scat}|}\frac{\partial E_\alpha^{scat^*}}{\partial x_\beta}\frac{1}{2|\mathbf{E}_{inc}|}\frac{\partial E_\gamma^{inc}}{\partial x_\delta}\times \\ \times\frac{1}{2|\mathbf{E}_{inc}|}\left(\frac{\partial E_x^{inc}}{\partial x} - \frac{\partial E_y^{inc}}{\partial y} - 2\frac{\partial E_z^{inc}}{\partial z}\right)\end{array}\right)_{surf} +$$

$$+ \sum_{\substack{\alpha,\beta,\chi,\eta \\ \alpha\neq\beta \\ \chi\neq\eta}} C_{V_{(s,p)}}\left[Q_{\alpha\beta}, Q_2, Q_{\chi\eta}\right]\left(\begin{array}{c}\frac{1}{2|\mathbf{E}_{scat}|}\frac{\partial E_\alpha^{scat^*}}{\partial x_\beta}\times \\ \times\frac{1}{2|\mathbf{E}_{inc}|}\left(\frac{\partial E_x^{inc}}{\partial x} - \frac{\partial E_y^{inc}}{\partial y} - 2\frac{\partial E_z^{inc}}{\partial z}\right)\frac{1}{2|\mathbf{E}_{inc}|}\frac{\partial E_\chi^{inc}}{\partial x_\eta}\end{array}\right)_{surf} +$$

$$+ \sum_{\substack{\gamma,\delta,\chi,\eta \\ \gamma\neq\delta \\ \chi\neq\eta}} C_{V_{(s,p)}}\left[Q_2, Q_{\gamma\delta}, Q_{\chi\eta}\right]\left(\begin{array}{c}\frac{1}{2|\mathbf{E}_{scat}|}\left(\frac{\partial E_x^{scat^*}}{\partial x} - \frac{\partial E_y^{scat^*}}{\partial y} - 2\frac{\partial E_z^{scat^*}}{\partial z}\right)\times \\ \times\frac{1}{2|\mathbf{E}_{inc}|}\frac{\partial E_\gamma^{inc}}{\partial x_\delta}\frac{1}{2|\overline{\mathbf{E}}_{inc}|}\frac{\partial E_\chi^{inc}}{\partial x_\eta}\end{array}\right)_{surf} +$$

$$+ \sum_{\substack{\alpha,\beta \\ \alpha\neq\beta}} C_{V_{(s,p)}}\left[Q_{\alpha\beta}, Q_2, Q_2\right]\left(\begin{array}{c}\frac{1}{2|\mathbf{E}_{scat}|}\frac{\partial E_\alpha^{scat^*}}{\partial x_\beta}\times \\ \times\frac{1}{4|\mathbf{E}_{inc}|^2}\left(\frac{\partial E_x^{inc}}{\partial x} - \frac{\partial E_y^{inc}}{\partial y} - 2\frac{\partial E_z^{inc}}{\partial z}\right)^2\end{array}\right)_{surf} +$$

$$+ \sum_{\substack{\gamma,\delta \\ \gamma \neq \delta}} C_{V_{(s,p)}}[Q_2, Q_{\gamma\delta}, Q_2] \left( \frac{1}{2|\mathbf{E}_{scat}|} \left( \frac{\partial E_x^{scat^*}}{\partial x} - \frac{\partial E_y^{scat^*}}{\partial y} - 2\frac{\partial E_z^{scat^*}}{\partial z} \right) \times \right.$$
$$\left. \times \frac{1}{2|\mathbf{E}_{inc}|} \frac{\partial E_\gamma^{inc}}{\partial x_\delta} \frac{1}{2|\mathbf{E}_{inc}|} \left( \frac{\partial E_x^{inc}}{\partial x} - \frac{\partial E_y^{inc}}{\partial y} - 2\frac{\partial E_z^{inc}}{\partial z} \right) \right)_{surf} +$$

$$+ \sum_{\substack{\chi,\eta \\ \chi \neq \eta}} C_{V_{(s,p)}}[Q_2, Q_2, Q_{\chi\eta}] \left( \frac{1}{2|\mathbf{E}_{scat}|} \left( \frac{\partial E_x^{scat^*}}{\partial x} - \frac{\partial E_y^{scat^*}}{\partial y} - 2\frac{\partial E_z^{scat^*}}{\partial z} \right) \times \right.$$
$$\left. \times \frac{1}{2|\mathbf{E}_{inc}|} \left( \frac{\partial E_x^{inc}}{\partial x} - \frac{\partial E_y^{inc}}{\partial y} - 2\frac{\partial E_z^{inc}}{\partial z} \right) \frac{1}{2|\mathbf{E}_{inc}|} \frac{\partial E_\chi^{inc}}{\partial x_\eta} \right)_{surf} +$$

$$+ C_{V_{(s,p)}}[Q_2, Q_2, Q_2] \left( \frac{1}{2|\mathbf{E}_{scat}|} \left( \frac{\partial E_x^{scat^*}}{\partial x} - \frac{\partial E_y^{scat^*}}{\partial y} - 2\frac{\partial E_z^{scat^*}}{\partial z} \right) \times \right.$$
$$\left. \times \frac{1}{4|\mathbf{E}_{inc}|^2} \left( \frac{\partial E_x^{inc}}{\partial x} - \frac{\partial E_y^{inc}}{\partial y} - 2\frac{\partial E_z^{inc}}{\partial z} \right)^2 \right)_{surf}$$

(8)

From (2-8) one can see, that all the scattering contributions depend only from dipole and minor quadrupole moments, that is associated with Electrodynamical forbiddance of the strong quadrupole light-molecule interaction.